**Version 1.**

**From Single to Multi Mode Lasing: The role of materials revealed in optical simulations.**

P.V. Shibaev[1*], A. Roslyak[1], P. Fessatidis[2]

[1]*Department of Physics, Fordham University, New York, NY 10458, USA*

[2]*Department of Physics, Boston college, Chestnut Hill, MA 02467, USA*

*corresponding author: P.V.Shibaev, shibpv@yahoo.com and shibayev@fordham.edu

# From Single to Multi Mode Lasing: The role of materials revealed in optical simulations.

Comparative study of thin film Cholesteric Liquid Crystal (CLC) lasers made from different materials and optically pumped by external solid state laser reveals a striking dependence of lasing behaviour (ranging from single mode at the edge of the selective reflection band to multimode lasing) on the morphology and microstructure of CLC films. The materials studied belong to two groups: low molar mass liquid crystals and polymers. It is shown that the orientation of individual chiral domains and fluctuations in helical pitch contribute significantly to the type of lasing displayed by the material. Different ways of preparing CLC cells that lead to predominantly one type of lasing are discussed. The importance of variations of the helical pitch and domain orientation in producing single and multimode lasing is justified by optical simulations (4x4 matrix method) of lasing in multi-layered samples.

**Introduction**

Light scattering in the periodic helical structure of CLC produces a stop band (selective reflection band - SRB) for the light with the same sense of polarization as the helicity of the CLC, which can be considered as a structure of twisted planar nematic planes with ordinary and extraordinary indices of refraction contributing to the birefringence $\Delta n = n_e - n_o$ [1]. In case of planar orientation of cholesteric planes and probing light beam directed along the helical axis the spectral position of the centre of the SRB is defined by the wavelength:

$$\lambda = \frac{P(n_o + n_e)}{2} \qquad (1)$$

where P is a helical pitch of chiral CLC structure.

Dye-doped and optically pumped CLCs has attracted the interest of scientific community for a long time [2,3], but the spectral attribution of lasing in CLCs to the band edge modes near the SRB was made only twenty years later [4]. This sparked renewed research interest to the band edge lasing in different types of CLCs [5-13], including thermotropic and lyotropic liquid CLCs confined between two flat glass substrates [5], responsive polymer CLCs of different nature [6], compositions of CLC with polymers [7,8]. The accounts of the research were reviewed in [9-13].

Depending on physical state of active media the micro lasers based on CLCs will be refereed in this article as liquid and polymer lasers. Liquid and polymer CLC lasers based on almost ideal planar structures were shown to display lasing at one or few modes near the band edge. The slope efficiency of these lasers is relatively high [10]. The birefringence of twisted nematic planes and their planarity determines the efficiency of planar chiral resonator [11,12]. Low threshold lasing from chiral droplets and colloids was also reported in Ref [13].

CLCs as one dimensional photonic band gap materials for circularly polarized light may also exhibit localized modes inside the band gap if periodicity of the helical structure is somehow disrupted. The possible use of this property for achieving low threshold lasing gave an additional boost to research in the area of solid CLCs, with defects specifically designed and prepared either by photopolymerization of monodomain liquid CLC samples, or by utilizing polymer networks with randomly distributed defects [14,15]. Lasing at specially created defects in otherwise planar cholesteric structures is often referred as lasing at localised or "defect modes" [15,16]. However, it is important to underscore that vitrification of melted CLC [14] as well as polymerization of CLC samples often results in the appearance of additional inhomogeneities (not necessarily "desirable"), for example, pitch variations across the sample. The extra defects can also be introduced in polymer stabilized CLCs or polymer dispersed liquid crystals in which the droplets of CLC are embedded in polymer matrix or domains of CLCs are separated by the elements of polymer networks [17]. Lasing modes can be excited not only at the spectral edge of the SRB band but also inside it [18-20]. The defects and inhomogeneities can be created in CLC matrix by different means, for example by adding nanoparticles [18], large polymer molecules [19] or by fast temperature changes of CLCs leading to the appearance of disoriented domains [20]. This type of lasing is referred as a random lasing [21], the type of lasing with emitted light forming a closed path due to multiple light scattering at optical inhomogeneities. Random lasing in many aspects is related to a light localization problem [22,23].

Interestingly, in most papers published on lasing only one type of lasing (single mode, lasing from defects or random lasing) is considered. It makes understanding transitions between different types of lasing in non-ideal planar samples made from different materials difficult.

The purpose of this paper is to fill this knowledge gap by considering experimental techniques of creating mono- and multi- domain samples with different types of disorder resulting in different types of lasing. We also present a computational model describing the transition between different lasing modes and show that lasing from the samples with different types of defects leads to multimode lasing either at the edge of the SRB or inside it.

**Experimental**

Some chemical structures of the monomer compound used to produce multidomain samples are shown in Figure 1. Merck mixture E63 used together with the chiral dopant CB15 in order to produce monodomain CLC with the SRB centred at 590nm with the spectral bandwidth about 60nm. The mixture is characterized by low viscosity and, therefore, it forms unstable defects structure.

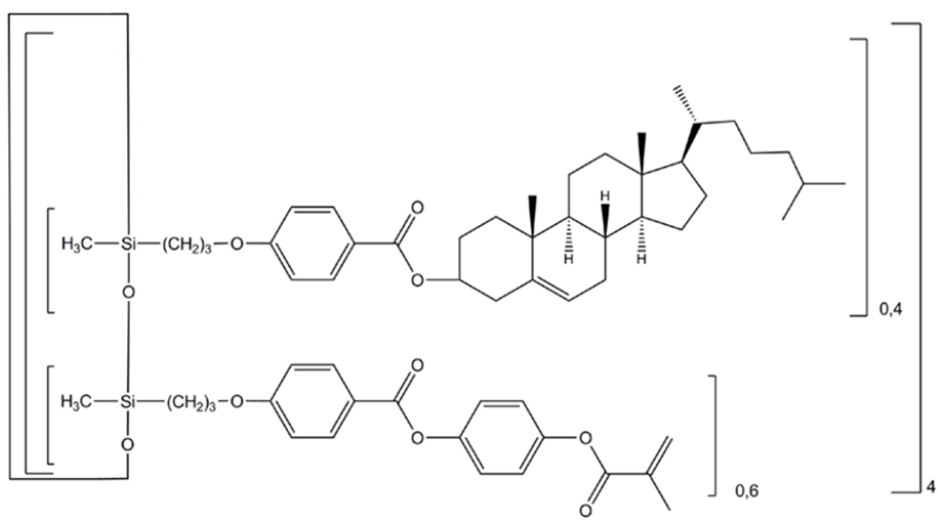

**Figure 1** Chemical structure of the monomer (percentage of chiral and non-chiral groups can vary)

Wacker monomers (Figure 1) were used in order to synthesize polymer samples with different numbers of defects and to study the effects related to multimode lasing. These monomers melt and form CLC at c.a. 70°C, they are much more viscous than liquid CLC samples (the estimation of viscosity by two plates method [7] indicates to at least three orders of magnitude difference). Thus, the oil streak and other defects appearing in the sample are more persistent and need more time to quench. These compounds are precursors for polymer CLCs and referred in the text as viscous CLCs.

Pyrromethane 597 was used as a lasing dye in both compositions in mass concentration of c.a. 0.2%. This dye has high quantum efficiency and relatively large spectral distance between absorption and emission bands. The liquid and viscous CLC compositions were optimized in order to get high frequency edge of the SRB close to the maximum of dye emission peak positioned at c.a. 580nm.

The following technique allowed the creation of high quality monodomain and multidomain samples.

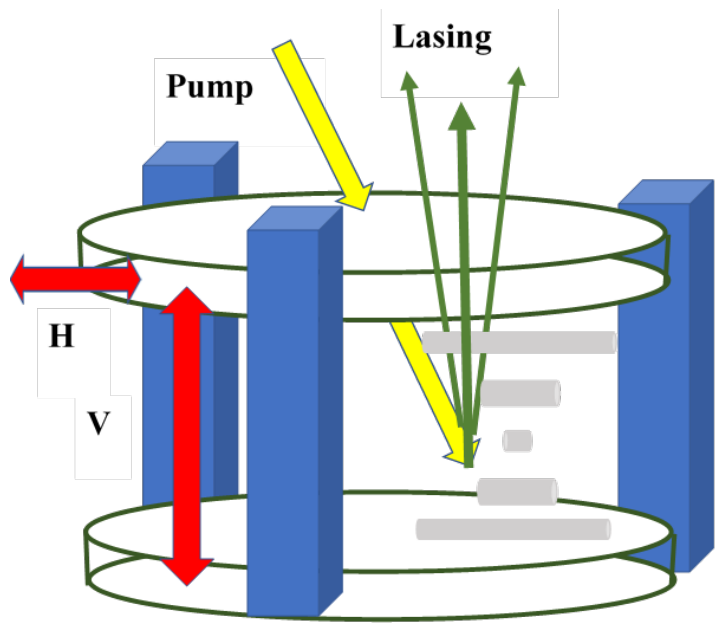

**Figure 2. Setup for creating planar liquid and polymer CLC samples**

Two glass plates with surface flatness of about one tenth of the green light wavelength ($\lambda$=550nm) were kept in a holder with a precise and controllable adjustment of the gap between the glass plates. This was achieved by three high precision screws with a rotation corresponding to 0.1 micron shift along the z axis of the setup (Figure 2). The additional screw H on the upper plate allowed for a horizontal motion of the upper plate in order to create inhomogeneous defects inside the sample. The parallelism of two glass plates was maintained visually by following the interference pattern between the glasses. In order to create planar orientation, the glasses were treated with polyimide following the standard annealing and rubbing procedure. The monodomains of different thickness are formed by injection of CLC between the plates with subsequent increase the distance between the plates while maintaining their parallelism. Large monodomains were formed when the distance between the plates was about 10 microns. With the increasing thickness of the sample the size of the monodomain decreased as defects appeared inside the sample due to the addition of CLC. For a sample with the thickness of 40 microns the typical monodomain size of liquid CLC was about 10mm (as determined by the distance between two oil streaks).

The same setup was used to create polymer samples from viscous CLCs. The size of the monodomains viewed from the above was typically smaller, about hundreds of microns. The whole system was heated to the appropriate temperature above 70º C and UV light was shined from both sides on the sample in order to induce polymerization and reduce the effect of helical pitch variation across the sample. Highly inhomogeneous samples could also be created by the aforementioned setup if one of the plates moved horizontally during sample preparation.

The lasing experiments were performed as follows. The beam from YAG laser hits the sample at an angle 20-30 degrees. The beam is focused on CLC by a long focal length lens allowing the formation of narrow and lengthy "neck". The spectra were recorded by Ocean Optics spectrometer collecting light in a relatively narrow cone of 5-10 degrees.

**Results and Discussion**

*Monodomain Samples*

The typical transmission and emission spectra of liquid monodomain CLC sample are shown in figure 3a. The oscillations of transmission near the band edge are clearly

a)                                                          b)

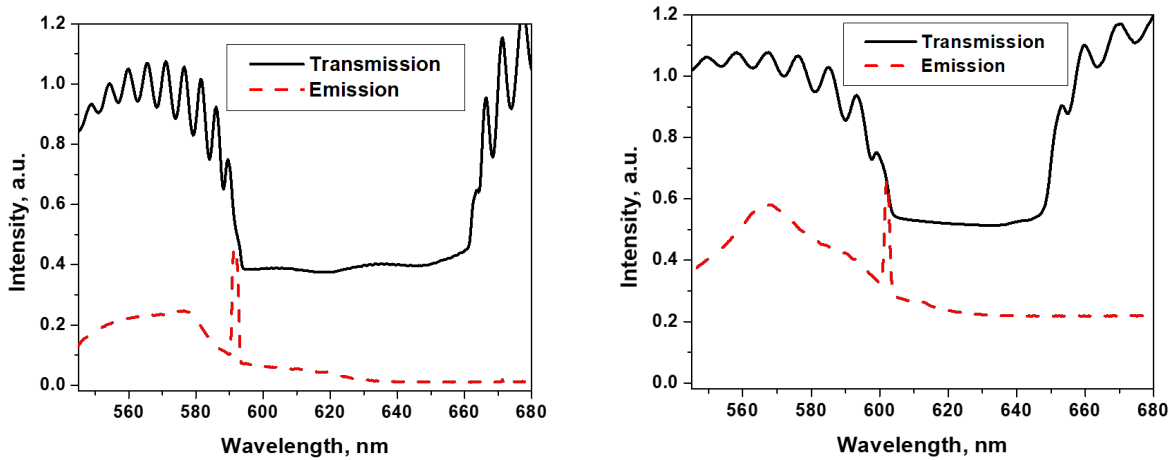

**Figure 3. Transmission and lasing from monodomain liquid and polymer samples. a) liquid CLC, b) polymer CLC.**

visible. In a stable monodomain sample lasing occurs at the first mode near the band edge. Often, when few lasing pulses hit the same spot of the sample a few more lasing modes can be excited (thermal effects lead to local reorganization of CLC structure, changes in helical pitch and planarity). The lasing threshold decreases with increasing thickness of the sample up to 35 microns and then stabilizes. The absolute value of lasing threshold in polymer samples compared to one in liquid samples is normally higher by a factor of 2-5, but this study mostly concentrates on positions of lasing modes with respect to the SRB in liquid and polymer CLC samples.

The transmission spectrum of polymer samples with moderate thickness is less featured with smaller and less distinctive band edge oscillations (Figure 3b) than liquid CLC samples of the same thickness. The lasing threshold in polymer samples is often higher and starts to saturate at a thickness of about 25 microns. Polymer samples prepared by irradiation of just one side of the sample display slight variation of reflected colour from two sides of the sample with blue colour shift on the side of the sample placed closer to UV light source (the shrinkage of polymer during polymerization results in shorter helical pitch). In thicker samples this effect is more pronounced.

Thus, polymerization of thin samples (with a thickness about 10 microns) results in monodomain samples with relatively uniform pitch distribution and distinctive band edge oscillations. However, polymerization of thick samples often introduces a helical pitch gradient and variations in domain orientations across the sample.

*Multidomain Samples*

Multidomain liquid samples can be easily created either by quick addition of pure CLC (or CLC mixed with nanoparticles) or by quick shift of the upper plate with respect to the lower glass (Figure 2). Examination of the samples under the microscope revealed that both methods resulted in the appearance of numerous oil streak defects in liquid samples and small penetrating domains with average size less than 1mm. In pure CLC samples the oil streak defects slowly disappeared, while in viscous CLC samples oil streak defects were stable for hours. The domains positioned close to the glass substrates retained their planarity, while domains positioned further from the glass substrate seemed to have slightly different colours. Lasing from the areas with oil streaks was suppressed due to the intense light scattering from individual oil streaks or their aggregates. Interestingly, the small horizontal disturbance leading to deformation of liquid CLC without creating stable oil streak resulted in lasing appearing at higher modes and increased lasing threshold. Further deformation leaded to the disappearance of narrow lasing lines replaced with broader amplified emission from the sample. Transmission spectra of areas with oil streaks are broad with transmission of incident light less than 50% that indicates higher light scattering.

Polymer multidomain samples, however, had a different morphology. The oil streaks were still present, but the structure of their conglomerates was washed out. The domains reflected different colours due to the variations in helical pitch and orientation of the individual domains. Transmission spectra are also broad with the sloping band edge. Lasing from these samples often display several peaks inside the SRB with a higher threshold than lasing in monodomain samples (Figure 4). The interesting feature of lasing spectrum is that sharp peaks appear on the top of otherwise broad emission spectrum and their position is maintained within the spectrum.

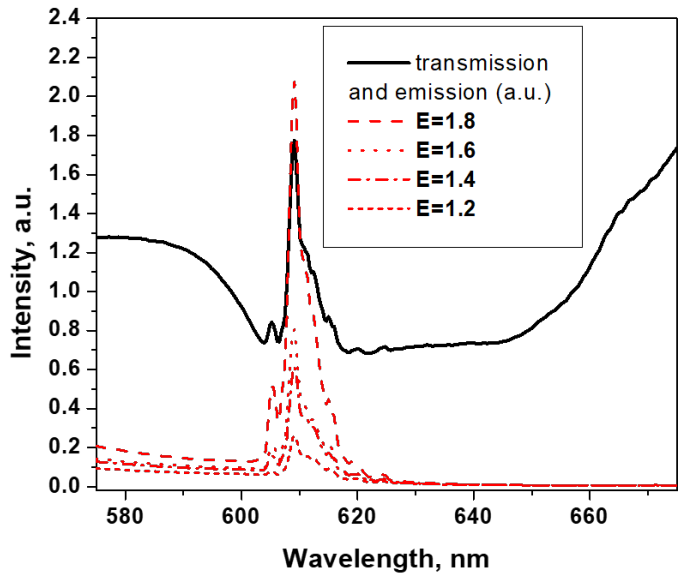

**Figure 4. Transmission and lasing from multidomain polymer sample**

The latter is often considered as a characteristic of random lasing when light scattering from randomly distributed scatterers creates a closed loop for light thus playing a role of resonator in classical lasing [20-22]. Random lasing was observed in cholesteric liquid crystals doped with nanoparticles [18]. It was also observed in polymer dispersed CLCs [17], where the lasing was attributed to numerous scattering events in the disordered material and lasing from highly disoriented droplets.

In order to understand whether the observed lasing behaviour (see Figures 3,4) can be understood in the framework of the multidomain model with distribution of different types of defects (disoriented domains in case of liquid samples and domains with helical pitch change in case of polymer samples) the optical modelling of lasing was conducted.

*Optical Modelling*

Optical calculations were performed in the framework of Berreman 4x4 matrix method [24] in order to clarify the nature of the transition from mono- to multi - mode lasing occurring in monodomain and multidomain liquid and polymer samples. The following model was employed in order to study lasing from monodomain samples and samples without the light scatterers. The planar CLC slabs with a thickness of about 10-15 helical pitches were separated by thin isotropic layers with the average refractive index equal to the average refractive index of the slab (in order to avoid effects related to the Fresnel reflection and appearance of localized modes originating from the disruption in refractive index).

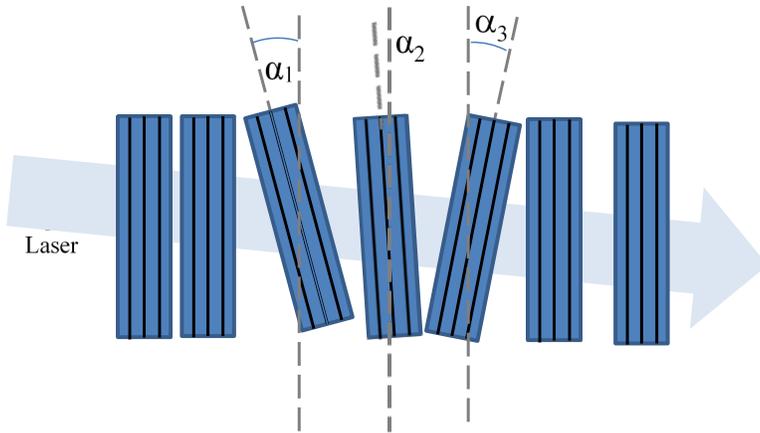

**Figure 5. Structural model of CLC sample used in optical calculations. The distances between slabs are exaggerated.**

The calculations were performed for a set of seven CLC slabs (Figure 5) four of which were fixed and three internal ones were allowed to disorient by angle $\alpha$ simulating the disorientation in inner domains in liquid and polymer samples with respect to the CLC surface. In the framework of Berreman model the disorientation essentially leads to a variation of the angle of incident light. Helical pitch was allowed to change in all slabs. In case of polymer models the helical pitch of two outer slabs was shorter in order to account for spectral blue shift of outer CLC layers occurring during polymerization.

The simulation of lasing was performed by introducing positive gain (imaginary part $\gamma$ of the refractive index, $\varepsilon=\varepsilon_0+i\gamma$) for the spectral interval corresponding to the emission of real dye. Increasing gain corresponds to higher pump energy in real experiments. However, no quantitative correspondence can be established between the two. The spectral position of lasing peaks with respect to the SRB was the major subject of this study. The shape of the emission band was approximated by a Lorentzian. The maximum of emission was at c.a. 575 nm and the width of emission band was about 60nm.

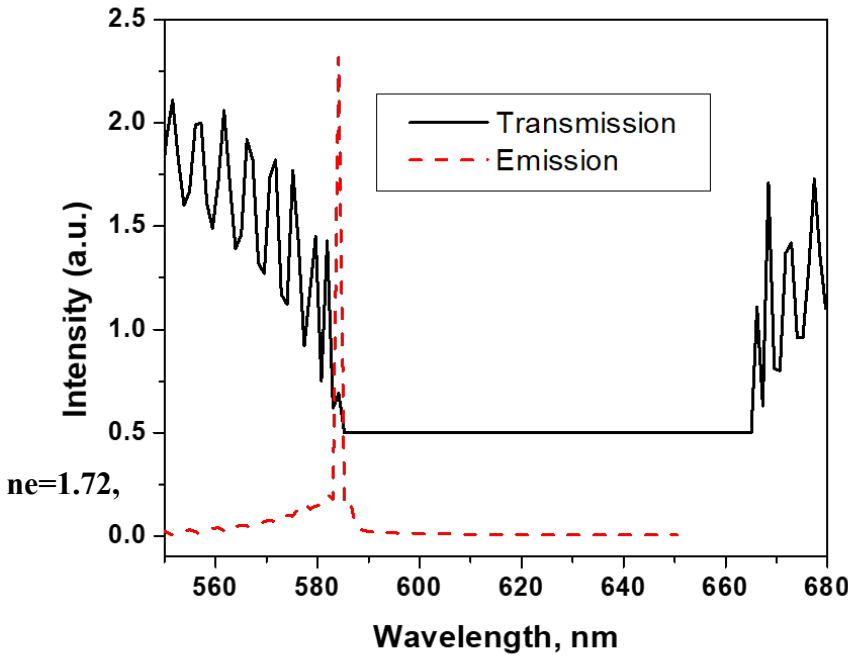

**Figure 6. Calculations of light transmission and lasing from monodomain sample with optical parameters reflecting experimental data: ne=1.72, no=1.54, P=400nm**

The validity of this model was checked by reproducing transmission optical spectrum of perfectly planar monodomain sample discussed in the beginning of the experimental section and lasing from this sample (Figure 6). All CLC slabs of this sample were planar and parallel to each other. The indexes of refraction were $n_e$=1.72 and $n_o$=1.54, the helical pitch P= 400nm. It can be seen that the correspondence between experimental transmission spectrum and the calculated one is very good. The simulation of lasing also shows the appearance of the peak at the first higher frequency mode ($\gamma$ =0.002).

Polymer samples can also be characterized by variations in helical pitch in individual domains. The birefringence of Wacker polymer (Wacker company's data) was $\Delta n=n_e-n_0=0.1$. The position and shape of the SRB allow to estimate the average helical pitch of the sample P=400nm, assuming the average refractive index $n_{av}$=1.55. These data were used in calculations in order to simulate the light transmission and lasing from these samples. The experimental transmission spectrum of thick samples is very broad due to the fluctuations in helical pitch, order parameter and domain orientations. So, the purpose of the modelling was not to reproduce the whole spectrum, but rather examine the role of changes in helical pitch and domain orientation occurring in polymer samples.

First, the ideal sample with cholesteric pitch P=400nm and δn=ne-n0=0.1was simulated as a reference for simulations of disorder. Results of optical properties calculations performed for ideal CLC sample (lasing angles were averaged over the cone of 8 degrees as in real experiments) are shown in Figure 7 (curves 1 and 2). The lasing peak was near the band edge; increasing the gain resulted in higher lasing intensity at the first mode near the SRB.

Second, the samples with varying helical pitch were studied. The internal slabs of the sample were assigned longer helical pitch. Outer slabs were assigned shorter helical pitch that changed from 380nm and increased to 420 nm in the centre of the sample. These numbers were chosen in accordance to experimental data showing the fluctuations in the spectral position of the SRB at different points of the sample. About forty transmission spectra were simulated for different incident angles, some of them showed sharp changes in transmission spectra over the spectral area of the SRB. Averaging over all these angles produced broad and featureless transmission spectra. The results of averaging over eight spectra are presented at Figure 7 (curves 3 and 4), they still contain drops in transmission at 590nm and 610nm. The lasing also occurred at 590nm and 610nm.

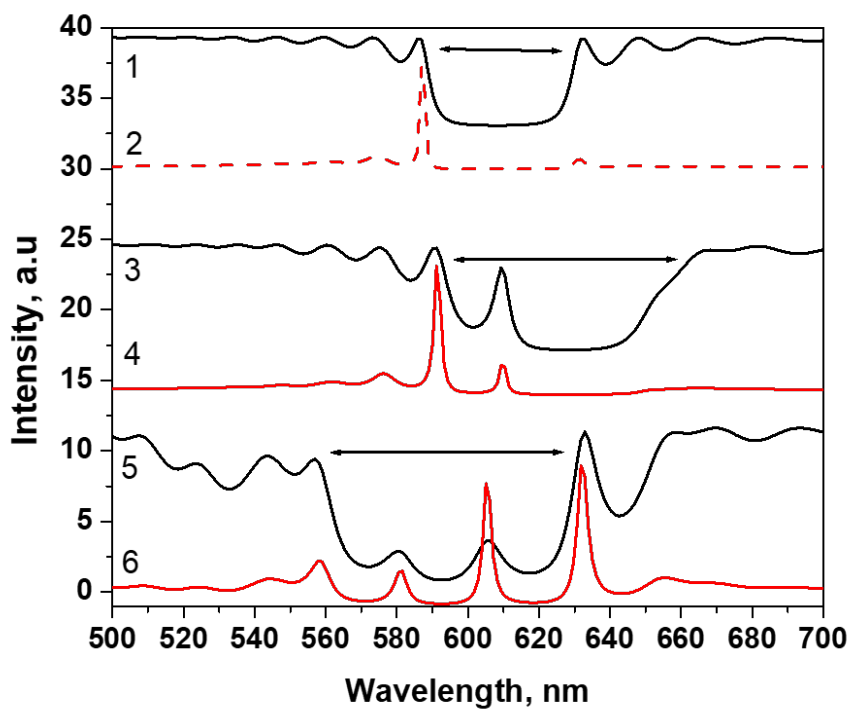

Figure 7 Results of optical calculations of three different structures: curves 1 and 2- ideal planar structure with P=400nm, curves 3 and 4 – planar structure with different helical pitches P (within 380nm<P<420nm), curves 5 and 6 – structure with varying helical pitches and orientations of slabs (within 20 degrees).

The results of optical calculations of samples with varying helical pitch and slabs orientations were shown in Figure 7 (curves 5 and 6). The angles α between internal domains (see Figure 5) were changing between 0 and 20 degrees. The transmission spectra and lasing peaks were also averaged inside the cone of lasing (over eight degrees). This model produced four lasing peaks lying at the different wavelengths, but inside the SRB at c.a. 560nm, 580nm, 605nm and 630nm.

It is important to note that in the studied model the light propagates along the straight line and results are averaged over different paths as it happens in real experiments. The model considers scattering effects only along straight paths. In spite of such a restriction the multiple lasing peaks are still produced in highly disordered samples within the SRB.

## Conclusions

Mono and multidomain samples of cholesteric liquid crystals were synthesized and their optical properties including optically pumped lasing were studied. Large liquid monodomain CLCs can be prepared as very thick samples (about 120 microns). Large and uniform polymer monodomain samples are difficult to prepare due to the shrinkage of the monomer during polymerization as well as appearance of inhomogeneities and defects. Single mode lasing occurred in monodomain samples and multimode lasing occurred in multidomain samples with disoriented domains and the presence of oil-streak defects. Optical simulations show that multimode lasing can be explained by introducing inhomogeneities ( out of plain deviations from planar orientation and helical pitch variations). The more inhomogeneous the samples, the more lasing peaks appear inside the SRB.

**Acknowledgements**

This work was in part supported by Fordham Fellowship Program provided sabbatical leave for PS

**References**

[1] de Gennes PG, Prost J. The physics of liquid crystals, Claredon Press, Oxford, 1998

[2] Kukhtarev NV. Cholesteric liquid crystal laser with distributed feedback. Sov J Quantum Electron. 1978;8(6):774-776.

[3] Ilchishin IP, Tikhonov EA, Tischenko VG, Shpak MT. Generation of a tunable radiation by impurity cholesteric liquid crystals. JETP Lett. 1980;32(1):24-27.

[4] Kopp VI, Fan B, Vithana HK, et al. Low-threshold lasing at the edge of a photonic stop band in cholesteric liquid crystals. Opt Lett. 1998;23(21):1707-1709.

[5] Shibaev P, Genack AZ, Narrowing of spontaneous emission and lasing in lyotropic and thermotropic liquid crystals. Liq Cryst. 2003;30:1365-1368.

[6] Shibaev P, Chiappetta D, Sanford RL, et al. Color changing cholesteric polymer films sensitive to amino acids. Macromolecules. 2006;39(12):3986-3992.

[7] Shibaev P, Crooker B, Manevich M, et al. Mechanically tunable microlasers based on highly viscous chiral liquid crystals. Appl Phys Lett. 2011;99(23):233302-233302-3.

[8] Shibaev PV, Rivera P, Teter D, et al. Color changing and lasing stretchable cholesteric films. Opt Exp. 2008;16(5):2965-2970.

[9] Palto SP, Blinov LM, Barnik MI, et al. Photonics of liquid-crystal structures: A review. Crystallogr Rep. 2011;56(4):622-649.

[10] Ford AD, Morris SM, Coles HJ. Photonics and lasing in liquid crystals. Mater Today. 2006;9:36-42.

[11] Dudok TH. Nastishin YA. Optically pumped mirrorless lasing. A Review. Part II. Lasing in photonic crystals and microcavities. Ukr J Phys Opt. 2014;15(2):47-67.

[12] Ortega J, Folcia CL, Etxebarria J. Upgrading the Performance of Cholesteric Liquid Crystal Lasers: Improvement Margins and Limitations. Materials. 2018;11(1):5.

[13] Muševič I. Liquid-crystal micro-photonics. Liq Cryst Rev. 2016;4(1):1-34.

[14] Shibaev PV, Kopp VI, Genack A, et al. Lasing from chiral photonic band gap materials based on cholesteric glasses. Liq Cryst. 2003;30:1391-1400.

[15] Vetrov S, Timofeev I, Shabanov V. Localized modes in chiral photonic structures. Phys Usp. 2020;63(1):33-56.

[16] Belyakov VA. Localized modes in optics of chiral liquid crystals. Mol Cryst Liq Cryst. 2015;612(1):81-97.

[17] Shibaev PV, Kopp VI, Genack A. Photonic Materials based on mixtures of cholesteric liquid crystals with polymers. J. Phys. Chem. B 2003; 107:6961-6964

[18] Chen SC, Lin JD, Lee CR, et al. Multi-wavelength laser tuning based on cholesteric liquid crystals with nanoparticles. J Phys D: Appl Phys. 2016;49:165102.

[19] Lu H, Xing J, Wei C, et al. Band-gap-tailored random laser. Photonics Res. 2018;6(5):390-395.

[20] Gao S, Wang J, Li W, et al. Low threshold random lasing in dye-doped and strongly disordered chiral liquid crystals. Photonics Res. 2020;8(5):642-647.

[21] Sapienza R. Determining random lasing action. Nature Reviews. 2019;1:690-695

[22] Genack AZ, Chabanov AA. Signatures of photon localization. J Phys A: Math Gen. 2005;38(49):10465-10488.

[23] Gevorgyan AH. Specific properties of light localisation in the cholesteric liquid crystal layer. The effect of layer thickness. Liq Cryst. 2020; 47: 1070-1077

[24] Berreman DW. Optics in stratified and anisotropic media: 4x4 matrix formulation. J Opt Soc Am. 1972;62(4):502-510.